\documentclass[a4paper,12pt]{iopart}
\usepackage[cp1250]{inputenc}
\usepackage[]{amssymb}
\usepackage[]{amsthm}
\usepackage[]{mathrsfs}
\usepackage[]{tensor}
\usepackage[]{eucal}
\usepackage[symbol]{footmisc}
\newcommand{\M}{\mathscr{M}}

\newcommand{\ii}{i}
\newcommand{\ip}[2]{\left<#1,#2\right>}
\newcommand{\hg}[1]{{*#1}}
\newcommand{\hh}{_\mathrm{H}}
\newcommand{\qqd}{\ , \quad}
\newcommand{\be}{\begin{equation}}
\newcommand{\ee}{\end{equation}}
\newcommand{\bea}{\begin{eqnarray}}
\newcommand{\eea}{\end{eqnarray}}
\newcommand{\0}{\nonumber}
\newcommand{\eH}{\buildrel H \over =}
\newcommand{\eB}{\buildrel \mathcal{B} \over =}

\newcommand{\prf}[1]{\noindent \textbf{Proof.} #1 \qed}

\begin{document}

\begin{flushright}
ZTF-12-03
\end{flushright}

\title[Killing Horizons as Equipotential Hypersurfaces]{Killing Horizons as Equipotential Hypersurfaces}

\bigskip

\author{Ivica Smoli\'c}

\address{Department of Physics, Faculty of Science, University of Zagreb, p.p.~331, HR-10002 Zagreb, Croatia}

\ead{ismolic@phy.hr}

\vspace{20pt}

\begin{abstract}
In this note we present a new proof that Killing horizons are equipotential hyper\-surfaces for the electric and the magnetic scalar potential, that makes no use of gravitational field equations or the assumption about the existence of bifurcation surface.
\end{abstract}



\vspace{20pt}

\section{Introduction}

It is well known that the electric and the magnetic scalar potential, $\Phi$ and $\Psi$, are constant over the horizons of stationary axisymmetric black hole solutions of Einstein-Maxwell field equations. Besides being a fascinating fact by itself, this piece of information is important in the formulation of the first law of black hole mechanics (conventional form of the ``potential-charge'' terms, $\Phi\hh\,\delta Q$ and $\Psi\hh\,\delta P$, is obtained by integrating out the electromagnetic scalar potentials), as well as in the proofs of the black hole uniqueness theorems (see e.g.~\cite{Heusler}, chapter 8). 

\bigskip

In order to prove a typical theorem in the physics of black holes, one has to choose the initial assumptions and there are usually two ideal paths: either to assume the validity of particular equations of motion (e.g.~Einstein's field equation) or the presence of certain symmetries (isometries and symmetries of additional fields in the spacetime). This choice depends on whether we want a theorem to be valid for a class of solutions of the chosen equations of motion (independently of the symmetries), or for a class of spacetimes sharing the same symmetries (independently of the equations of motion). Problems in practice can get especially tricky, so that we'll have to compromise and reach for both types of assumptions. For example, Carter has proved in \cite{C73} that the electric and the magnetic scalar potential are constant over the black hole horizon, using symmetries of the spacetime \emph{and} Einstein-Maxwell field equations. This proof has been repeated in a multitude of modern references, e.g.~\cite{C89,Heusler,H98lrr,FN,GaoWald01}, in essentially unaltered form. One could still ask whether it is possible to exploit the symmetries a little bit further in order to avoid some of the field equations.

\bigskip

For those theorems which establish the constancy of some quantity $\mathscr{Q}$ over the black hole horizon, there is a third, usually more elegant path. Suppose we have a Killing horizon $H[\xi]$, generated by the Killing vector field $\xi^a$, with the bifurcation surface $\mathcal{B}$, consisting of points on which the Killing vector field is vanishing. The assumption about the presence of bifurcation surface is not particularly strong since R\'acz and Wald have shown \cite{RW92} that one can find a \emph{local} extension of a neighborhood of a regular (nondegenerate) Killing horizon which is a proper subset of a regular bifurcate Killing horizon in the extended spacetime and this extension can be made \emph{global} under certain weak conditions \cite{RW95} which are, for example, automatically fulfilled in a circular spacetime. The idea in this approach is to prove that $\mathscr{Q}$ is constant over the bifurcation surface $\mathcal{B}$ and along the orbits of the Killing vector $\xi^a$, from where it follows that $\mathscr{Q}$ is constant over the whole horizon $H[\xi]$. The first such example was given by Kay and Wald \cite{KayWald91} for the zeroth law of black hole mechanics, the constancy of the surface gravity $\kappa$. More recently, Gao in \cite{Gao03} has proved that the electric scalar potential $\Phi$ is constant over the bifurcate Killing horizon, using the fact that $\Phi$ is identically zero on the bifurcation surface and that it remains constant along the orbits of the Killing vector (at least as long as the electromagnetic field doesn't become singular somewhere on the horizon). The third path, however, has one drawback: degenerate (extremal) black holes don't possess a bifurcation surface. This doesn't represent an obstacle for the zeroth law of black hole mechanics, since the surface gravity $\kappa$ is identically zero on the horizon in the extremal case, but this is \emph{a priori} not the case with the electromagnetic potentials.

\bigskip

In order to fill this gap in literature we shall present a simple proof which makes use of somewhat stronger symmetry condition on the electromagnetic field (circularity of the electric current), but is independent of the gravitational field equations or the presence of bifurcation surface. In section 2 we review Carter's and Gao's proofs with some technical remarks. In section 3 we present the new proof that Killing horizons are equipotential hypersurfaces for the electromagnetic scalar potentials. In the final section we make the concluding remarks.

\bigskip

Let us briefly comment on the notation and the conventions in the paper. We use equalities $\eH$ and $\eB$ to indicate the validity of the equation on, respectfully, the Killing horizon $H[\xi]$ and the bifurcation surface $\mathcal{B}$. Although the equations are written in the ``indexless'' language of differential forms, at several places we employ abstract index notation in order to clarify the nature of some objects. The inner product between vectors and 1-forms is denoted by $\ip{X}{Y} \equiv X_a Y^a$ and for a 2-form $\omega_{ab}$ we have $2\ip{\omega}{\omega} = \omega_{ab}\,\omega^{ab}$. Also, we adopt the natural system of units, $c = G = \epsilon_0 = 1$.

\vspace{20pt}

\section{Two traditional proofs }

Throughout the paper we assume that the 4-dimensional spacetime $(\M,g_{ab})$ has the Killing horizon $H[\xi]$, generated by the Killing vector field $\xi^a$, and the electromagnetic field described by the field strength 2-form $F_{ab}$, which is a solution of Maxwell's equations
\be\label{eq:Maxwell}
dF = 0 \qqd d\hg{F} = \hg{j}
\ee
and which is invariant under the action of the Killing vector field $\xi^a$,
\be\label{eq:symmF}
\pounds_\xi F = 0
\ee
Since the Lie derivative with respect to a Killing vector and the Hodge dual commute, it follows immediately that $\pounds_\xi\hg{F} = 0$. The electric and the magnetic field 1-forms $E_a$ and $B_a$ measured by an observer with 4-velocity $u^a$ are given by
\be
E = -\ii_u F \qqd B = \ii_u \hg{F}
\ee
In the case of stationary spacetimes, such as the Kerr-Newman black hole solution, there is a special class of observers, usually referred to as \emph{stationary observers} (see \cite{MTW}, chapter 33.4), whose 4-velocity $u^a$ is proportional to the Killing vector $k^a + \Omega\,m^a$ (cf.~section 3), with constant angular velocity $\Omega$. The normalization condition $\ip{u}{u} = -1$ restricts the range of $\Omega$ for which these observers can be timelike. However, at the horizon this range is reduced to the single value $\Omega\hh$ and the Killing vector $k^a + \Omega\hh m^a$ becomes null, so that it's not possible to properly normalize timelike 4-velocity $u^a$. In other words, there are no timelike stationary observers at the black hole horizon.

\bigskip

The natural way to deal with this was proposed by Carter in \cite{C73}: one can formally introduce the electric and the magnetic field which are well behaved at the horizon $H[\xi]$ by contracting $F$ and $\hg{F}$ with the Killing vector $\xi^a$ itself,
\be\label{eq:defEB}
E \equiv -\ii_\xi F \qqd B \equiv \ii_\xi \hg{F}
\ee
The symmetry of the electromagnetic field (\ref{eq:symmF}) and the Maxwell's equations (\ref{eq:Maxwell}) imply that $E_a$ is a closed form, and the same is true for $B_a$ in the vacuum case ($j = 0$),
\bea
dE = -d\ii_\xi F = -\pounds_\xi F + \ii_\xi dF = 0 \0\\
dB = d\ii_\xi \hg{F} = \pounds_\xi \hg{F} - \ii_\xi d\hg{F} = 0
\eea
Hence, using Poincar\'e's lemma one can conclude that, at least locally, there exist the electric and the magnetic scalar potentials, $\Phi$ and $\Psi$, such that
\be
E = d\Phi \qqd B = d\Psi
\ee
It follows directly from these definitions that the electric and the magnetic scalar potentials are constant along the orbits of the Killing vector $\xi^a$, $\pounds_\xi \Phi = 0 = \pounds_\xi \Psi$. Let us now review Carter's original proof.

\bigskip

\noindent
\textbf{Theorem 1.} \textit{Let $(\M,g_{ab},F_{ab})$ be a solution of Einstein-Maxwell equations with Killing horizon $H[\xi]$, generated by the Killing vector field $\xi^a$, and electromagnetic field $F_{ab}$, invariant under the action of the Killing vector field $\xi^a$ and nonsingular on $H[\xi]$. Then the electric and the magnetic scalar potential, $\Phi$ and $\Psi$, are constant over the Killing horizon $H[\xi]$.}

\bigskip

\prf{ 
General properties of the Killing horizons (see e.g.~Proposition 6.15 in \cite{Heusler}) imply that
\be
R(\xi,\xi) \eH 0
\ee
and it is straightforward to check the validity of two following equalities,
\be\label{eq:EE-BB}
\ip{E}{E} - \ip{B}{B} = \ip{\xi}{\xi}\ip{F}{F}
\ee
\be\label{eq:EE+BB}
\ip{E}{E} + \ip{B}{B} = 8\pi T(\xi,\xi)
\ee
where $T_{ab}$ is the electromagnetic energy-momentum tensor. The electro\-magnetic 2-form $F_{ab}$ is by assumption nonsingular on the horizon $H[\xi]$, so that $\ip{\xi}{\xi} \eH 0$ with (\ref{eq:EE-BB}) implies $\ip{E}{E} \eH \ip{B}{B}$. Furthermore, using Einstein's field equation and (\ref{eq:EE+BB}) we have
\be
0 \eH R(\xi,\xi) = 8\pi T(\xi,\xi) \eH 2\ip{E}{E}
\ee
and thus
\be
\ip{E}{E} \eH 0 \qqd \ip{B}{B} \eH 0
\ee
From $\ii_\xi E = 0$ and $\ii_\xi B = 0$ it follows that $E_a$ and $B_a$ are proportional to $\xi_a$ at the horizon $H[\xi]$,
\be
E \eH 4\pi\sigma_\mathrm{E}\,\xi \qqd B \eH 4\pi\sigma_\mathrm{B}\,\xi
\ee
where $\sigma_\mathrm{E}$ and $\sigma_\mathrm{B}$ are some functions\footnote{Conventional ``$4\pi$'' prefactors are adopted from the membrane paradigm, where these functions are related to the electric and the magnetic surface charge densities of the stretched horizon (for a comprehensive review see \cite{TPM}).}. This allows us to conclude that for any vector $t^a$ tangent to the horizon $H[\xi]$ we have
\be\label{eq:lietphi}
\pounds_t \Phi = \ii_t d\Phi = \ii_t E \eH 4\pi\sigma_\mathrm{E}\,\ii_t\xi \eH 0
\ee
and analogously $\pounds_t \Psi \eH 0$. Hence, $\Phi$ and $\Psi$ are constant over the $H[\xi]$.
}

\vspace{20pt}

Next we present Gao's proof in an expanded and somewhat modified form, so that it includes the magnetic scalar potential and avoids unnecessary use of the gauge field $A_a$, introduced through $F = dA$.

\bigskip

\noindent
\textbf{Theorem 2.} \textit{Let $\xi^a$ be a Killing vector field of a spacetime $(\M,g_{ab},F_{ab})$ with bifurcate Killing horizon $H[\xi]$ and electromagnetic field $F_{ab}$ which is nonsingular on $H[\xi]$ and which is invariant under the action of the Killing vector field $\xi^a$. Then the electric and the magnetic scalar potential, $\Phi$ and $\Psi$, are constant over the horizon $H[\xi]$.}

\bigskip

\prf{
From the definitions of electric and magnetic field (\ref{eq:defEB}), the fact that $\xi^a \eB 0$ and the assumption that $F_{ab}$ is nonsingular on $H[\xi]$ it follows that
\be
0 \eB E = d\Phi \quad \textrm{and} \quad 0 \eB B = d\Psi
\ee
So, $\Phi$ and $\Psi$ are constant over the bifurcation surface $\mathcal{B}$. Using the fact that $\Phi$ and $\Psi$ are constant along the orbits of $\xi^a$ we conclude that they remain constant on each connected component of the bifurcate Killing horizon $H[\xi]$.
}

\vspace{20pt}

\section{A new proof}

In this section we turn our attention to the stationary axisymmetric spacetimes $(\M,g_{ab})$ with the corresponding commuting Killing vectors $k^a$ and $m^a$ (the latter is supposed to be the axial Killing vector with closed orbits). It is convenient to introduce the \emph{Killing 2-form} $\rho_{ab}$,
\be
\rho \equiv k \wedge m
\ee
We assume that inside this spacetime there is a Killing horizon $H[\ell]$, generated by the Killing vector
\be
\ell^a = k^a + \Omega\hh m^a \qqd \ip{\ell}{\ell} \eH 0
\ee
The rigidity theorems, either in weak or strong version (see e.g.~\cite{Heusler}, chapter 6), prove that $\Omega\hh$, the ``angular velocity of the horizon'', assumes a constant value over the horizon $H[\ell]$. Again, the electromagnetic field is described by the 2-form $F_{ab}$, which is a solution of Maxwell's equations (\ref{eq:Maxwell}). We say that the electromagnetic field is stationary if it is invariant under the action of the Killing vector field $k^a$, $\pounds_k F = 0$, and axisymmetric if it is invariant under the action of the Killing vector field $m^a$, $\pounds_m F = 0$. We shall be particularly interested in the electric current $j_a$ which satisfies the \emph{circularity condition}
\be\label{eq:circj}
\rho \wedge j = 0
\ee
This is trivially true in the vacuum case, otherwise it's telling us that the only nonvanishing components of the current are those parallel to the Killing vectors. Assuming that we have the stationary axisymmetric electromagnetic field and the circular current $j_a$ (\ref{eq:circj}), Carter has shown \cite{C73} (see also \cite{Heusler}, Proposition 5.6) that in every region of stationary axisymmetric spacetime intersecting the rotation axis, Maxwell's equations imply
\be\label{eq:kmF}
\ii_m\ii_k F = F(k,m) = 0 \qqd \ii_m\ii_k \hg{F} = \hg{F}(k,m) = 0
\ee
The second of these two equations is equivalent to the electromagnetic circularity condition
\be\label{eq:Fcirc}
\rho \wedge F = 0
\ee
Furthermore, we adopt conventional notation for the inner products between the Killing vectors $k^a$ and $m^a$,
\be
-V \equiv \ip{k}{k} \qqd X \equiv \ip{m}{m} \qqd W \equiv \ip{k}{m}
\ee
We assume the absence of the closed causal curves in the exterior of the horizon, so that $X \ge 0$, with the equality holding only on the rotation axis (where $m^a$ vanishes). The inner products with the Killing horizon generator $\ell^a$ are given by
\be\label{eq:contracts}
\ip{k}{\ell} = -V + \Omega\hh W \qqd \ip{m}{\ell} = W + \Omega\hh X
\ee
Since $\ip{k}{\ell} \eH 0$ and $\ip{m}{\ell} \eH 0$, we have
\be\label{eq:omegahor}
\Omega\hh \eH \frac{V}{W} \eH -\frac{W}{X}
\ee
Using all these elements we present a new argument for the ``equipotential nature'' of the Killing horizons.

\vspace{20pt}

\noindent
\textbf{Theorem 3.} \textit{Let $(\M,g_{ab},F_{ab})$ be a stationary axisymmetric spacetime containing a Killing horizon $H[\ell]$ and a stationary axisymmetric electro\-magnetic field $F_{ab}$, nonsin\-gular on $H[\ell]$, which is a solution of Maxwell's equations with the electric current $j_a$. If the current satisfies the circularity condition, $\rho \wedge j = 0$, then the electric scalar potential $\Phi$ is constant on each connected component of $H[\ell]$. In the vacuum case, $j = 0$, the magnetic scalar potential $\Psi$ is constant on each connected component of $H[\ell]$. If the required conditions for the current are satisfied at least on some open set $\mathcal{O}$ which intersects the horizon $H[\ell]$, the conclusions remain valid on the subset $\mathcal{O} \cap H[\ell]$.}

\bigskip

\prf{
In what follows we shall frequently use the following two identities,
\be\label{eq:wedgecontract}
\ii_X (\alpha_p \wedge \beta_q) = (\ii_X \alpha_p) \wedge \beta_q + (-1)^p\,\alpha_p \wedge (\ii_X \beta_q)
\ee
\be\label{eq:iplin}
\ii_{a X + b Y} = a\ii_X + b\ii_Y
\ee
valid for any $p$-form $\alpha_p$, $q$-form $\beta_q$, vectors $X^a, Y^a \in TM$ and scalar fields $a$ and $b$. First we note that
\bea\label{eq:ctr1}
\ii_k \rho & = -Vm - Wk \eH -W\ell \0\\
\ii_m \rho & = Wm - Xk \eH -X\ell \\
\ii_k \hg{\rho} & = \ii_m \hg{\rho} = 0 \qqd \ii_m\ii_k \rho \eH 0 \0
\eea
Using (\ref{eq:wedgecontract}), (\ref{eq:iplin}), (\ref{eq:kmF}) and (\ref{eq:ctr1}), it is straightforward to check the following equalities,
\bea\label{eq:ctr2}
\ii_m \ii_k (\rho\wedge F) & \eH W\ell\wedge \ii_m F - X \ell\wedge \ii_k F \eH X \ell\wedge E \0\\
\ii_m \ii_k (\rho\wedge \hg{F}) & \eH W\ell\wedge \ii_m \hg{F} - X \ell\wedge \ii_k \hg{F} \eH -X \ell\wedge B \\
\ii_m\ii_k (\hg{\rho}\wedge F) & = 0 \0
\eea
with $E = -\ii_\ell F$ and $B = \ii_\ell \hg{F}$. The central idea of this proof is to contract equation (\ref{eq:Fcirc}) and the identity (see equation (7.179) in \cite{Nakahara})
\be
\rho\wedge\hg{F} = F \wedge \hg{\rho}
\ee
with $\ii_m\ii_k$. Using (\ref{eq:ctr2}) one gets
\be
0 \eH X\ell\wedge E = X\ell\wedge d\Phi \quad \textrm{and} \quad 0 \eH X\ell\wedge B = X\ell\wedge d\Psi
\ee
From these two equations it follows that at any point of $H[\ell]$, $d\Phi$ and $d\Psi$ are either zero or proportional to $\ell_a$ except possibly at points where $X = 0$ (note that at the points where $\ell = 0$ we have $d\Phi = 0$ and $d\Psi = 0$ by definition). Therefore, for any vector $t^a$ tangent to $H[\ell]$, $\ii_t d\Phi \eH 0$ and $\ii_t d\Psi \eH 0$, so that $\Phi$ and $\Psi$ are constant over $H[\ell]$ except possibly at points where rotation axis intersect the horizon $H[\ell]$. Since the electromagnetic field 2-form $F_{ab}$ is nonsingular on the horizon $H[\ell]$ we know that potentials are continuous on $H[\ell]$, and thus the conclusion about the constancy can be extended to all points of the horizon.
}

\vspace{20pt}

\section{Final remarks}

We have demonstrated that in order to prove that the electric scalar potential $\Phi$ and the magnetic scalar potential $\Psi$ assume constant values over the Killing horizon, the circularity of the electric current (\ref{eq:circj}) is enough to circumvent the gravitational field equations and avoid the additional assumption about the presence of bifurcation surface. It is interesting to note that the circularity of the electric current has already been used in the early analysis of Carter \cite{C73}, however, it hasn't been noticed that the same can be utilized to obtain a more general statement. It remains to be seen to what extent can all of these results be generalized in more complicated situations, such as those with nonsymmetrical fields. In the case of non-Abelian (Yang-Mills) fields, the definitions of analogous scalar potentials become more subtle, and their constancy over the horizon is usually used as a part of gauge fixing (see e.g.~\cite{Gao03} and references therein).

\vspace{20pt}

\ack

This work was supported by the Croatian Ministry of Science, Education and Sport under the contract no. 119-0982930-1016.

\vspace{25pt}

\section*{References}

\bibliographystyle{unsrt}
\bibliography{nEMKill}

\end{document}